\documentclass[twocolumn,amsmath,amssymb,aps,longbibliography]{revtex4}
\usepackage{graphicx}
\usepackage{amsmath,amssymb}
\usepackage{mathrsfs}
\usepackage{color}
\usepackage{afterpage}
\usepackage[version=3]{mhchem}
\usepackage{natbib}
\usepackage{soul}
\usepackage[caption=false]{subfig}
\usepackage{array}
\usepackage{multirow}
\usepackage[colorlinks, citecolor=blue,urlcolor=blue, linkcolor=blue, bookmarks=false]{hyperref}
\hypersetup{colorlinks=true , citecolor=blue, urlcolor=blue, linkcolor=blue}
\usepackage{braket}
\usepackage{bm}

\begin{document}

\title{Unlocking reversible and nonvolatile anomalous valley Hall control through multiferroic van der Waals heterostructures}

\author{Ankita Phutela\footnote{ankita@physics.iitd.ac.in}, Saswata Bhattacharya\footnote{saswata@physics.iitd.ac.in}} 
\affiliation{Department of Physics, Indian Institute of Technology Delhi, New Delhi 110016, India}
\begin{abstract}
\noindent 

Achieving external control over the anomalous valley Hall (AVH) effect is essential for advancing valleytronic applications. However, many of the existing approaches suffer from limitations such as irreversibility or volatility. In this work, we propose a general strategy for enabling nonvolatile electrical tuning of the AVH effect by utilizing multiferroic van der Waals heterostructures. Using first-principles density functional theory calculations, we demonstrate that a heterostructure composed of a ferromagnetic monolayer VSSe and a ferroelectric monolayer Al$_2$S$_3$ permits fine control of valley transport properties. The AVH response in VSSe can be reversibly and nonvolatility switched by reversing the polarization of Al$_2$S$_3$ via an applied electric field. This ferroelectric mechanism ensures a stable valley state even without continuous energy input. Furthermore, the valley polarization can also be inverted through the same polarization switching process, providing a dual degree of control over valley-dependent phenomena. These findings establish a promising pathway toward intrinsically switchable and energy-efficient valleytronic devices.
 
\end{abstract}
\maketitle

\section{Introduction}
The quest for novel quantum degrees of freedom and robust control mechanisms is driving the development of next-generation nanoelectronic devices. Among these, the valley degree of freedom, an energy extremum in the band dispersion of two-dimensional (2D) materials, has emerged as a promising information carrier. These binary valley states, found in energy-degenerate but inequivalent valleys at the Brillouin zone corners, form the foundation for valleytronic devices aimed at information storage and logical operations ~\cite{schaibley2016valleytronics, li2013valley, xiao2007valley, zeng2012valley}. Pioneering studies on 2D graphene and transition metal disulfides have revealed intriguing phenomena. These include the valley Hall effect and valley-dependent optical selection rules, paving the way for theoretical predictions and experimental realizations of applications such as valley spin valves and filters ~\cite{kioseoglou2012valley, kareekunnan2019magnetic, gunlycke2011valley, zhang2021valley}.

However, the widespread application of valley signals and the advancement of valleytronics for binary information encoding are often hindered by valley degeneracy. Various strategies have been proposed to overcome this by breaking time-reversal symmetry, including external electromagnetic fields, optical pumping, magnetic doping, or magnetic proximity effects ~\cite{zhou2021anomalous, mak2012control, cai2013valley, peng2018magnetic, sun2022electrically, guo2021electrically}. Unfortunately, many of these methods are either volatile, difficult to control, introduce undesirable impurity scattering, or yield only minuscule valley splitting, making them impractical for real-world applications ~\cite{mak2012control,guo2021electrically}. Crucially, achieving significant valley splitting through external means or magnetic substrates often presents experimental challenges. Thus, the central challenge for practical valleytronic devices lies in developing tunable and nonvolatile valley splitting.

To address this, the search for materials exhibiting intrinsic valley splitting is paramount. Inspired by the magnetic proximity effect, the coupled interplay of an intrinsic magnetic exchange field with spin-orbit coupling (SOC) is a highly desirable route to achieve nonvolatile intrinsic valley splitting. The breaking of both spatial inversion and time-reversal asymmetries in such systems leads to a nonzero valley magnetic moment and Berry curvature with unequal magnitudes and opposite signs, resulting in discernible valley splitting. Consequently, when a transverse voltage is applied, Bloch electrons acquire anomalous velocities, generating the anomalous valley Hall (AVH) effect. The coexistence of intrinsic magnetic exchange fields and SOC in ferromagnetic semiconductors makes them a focal point for research into nonvolatile valley splitting and the AVH effect. While explorations have spanned various 2D materials, including graphene, transition metal disulfides, and intrinsic ferrovalley materials like FeCl$_2$, GdI$_2$ and NdX$_2$ ~\cite{hu2016prediction, cheng2021prediction, zang2021ferrovalley}. Recent advancements have focused on heterostructures, such as electrical tuning of the AVH effect in antiferrovalley bilayer VSe$_2$, ferroelectric switching in inversion-symmetric bilayer T-FeCl$_2$ structures ~\cite{zhang2022spontaneous}, stacking-dependent AVH behavior in MnPSe$_3$/Sc$_2$CO$_2$ ~\cite{du2022anomalous}, and multiferroic systems like CrOX/In$_2$Se$_3$ ~\cite{sun2022reversible}, AgBiP$_2$S$_6$/CrBr$_3$ ~\cite{zhang2023nonvolatile} and WSe$_2$/VSe$_2$ ~\cite{marfoua2022reversal}.

Despite these advances, achieving reversible and nonvolatile control over valley and spin degrees of freedom in a unified material platform remains a significant challenge. In this work, we propose a concrete realization of such control using multiferroic van der Waals heterostructures. Specifically, we demonstrate through first-principles calculations that a heterostructure composed of ferromagnetic monolayer VSSe stacked on a ferroelectric monolayer Al$_2$S$_3$ enables precise, reversible modulation of the AVH response. By switching the ferroelectric polarization direction in Al$_2$S$_3$ via a transient electric field, the intrinsic valley polarization in VSSe can be flipped between the valence and conduction bands. This switch leads to an inversion of the AVH Hall voltage sign, providing a platform for electrically controllable, nonvolatile valleytronic functionalities. Our findings present a viable pathway toward practical and energy-efficient valley-based electronics.

\section{Computational Methods}
The calculations are performed at density functional theory (DFT)~\cite{hohenberg1964inhomogeneous,kohn1965self} level with the projector augmented wave (PAW)~\cite{kresse1999ultrasoft,blochl1994projector} method implemented in Vienna \textit{ab initio} Simulation Package (VASP)~\cite{kresse1996efficient} code.
The exchange-correlation interactions are treated by generalized gradient approximation (GGA) in the form of the Perdew-Burke-Ernzerhof (PBE) functional ~\cite{perdew1996generalized}. The cutoff energy of 520 eV is used for the plane wave basis set such that the total energy calculations are converged within 10$^{-5}$ eV. The vacuum in \textit{z}-direction is set to 30 \AA\ to avoid artificial interactions caused by the periodic boundary conditions. The 11$\times$11$\times$1 \textit{k}-grid is used to sample the 2D Brillouin zone. All the structural parameters are fully optimized until the Hellmann-Feynman forces are smaller than 1 meV/\AA.  A dipole correction ~\cite{neugebauer1992adsorbate} was considered. The interlayer van der Waals interaction was described by the DFT-D2 ~\cite{grimme2006semiempirical} method. Hybrid functional Heyd-Scuseria Ernzerhof (HSE06) employed to obtain accurate band structures ~\cite{heyd2003hybrid}.  The Berry curvatures are computed directly from the calculated wave functions using Fukui's method, as implemented in the VASPBERRY code ~\cite{kim2016competing,kim2022circular}.

\section{RESULTS AND DISCUSSION}
The advent of Janus materials, has opened new avenues in materials science due to their unique properties arising from broken out-of-plane mirror symmetry. The successful experimental synthesis of Janus MoSSe via modified chemical vapor deposition (CVD) methods has validated their practical realization ~\cite{cheng2013stable,lu2017janus}. This exciting progress, particularly the experimental synthesis of 2D ferrovalley materials like the 2H-VSe$_2$ monolayer, paves the way for a new class of Janus structures with intrinsic ferromagnetic semiconductor properties ~\cite{wang2021ferromagnetism}. Notably, the Janus 2H-VSSe monolayer was first predicted by Zhang et al., who thoroughly investigated its piezoelectricity, ferroelasticity, and valley polarization ~\cite{zhang2019first}. Given the success with similar materials, it is highly anticipated that ordered Janus 2H-VSSe can also be grown using modified CVD techniques. This prospect underscores the significance of gaining deeper insights into the intrinsic physical properties of the Janus 2H-VSSe monolayer. Our investigations reveal that VSSe is a ferromagnetic semiconductor exhibiting strong SOC. The loss of mirror symmetry in VSSe enables bidirectional modulation of the band gap simply by altering the direction of an applied electric field. With the simultaneous breaking of time-reversal, inversion, and mirror symmetries, VSSe promises enhanced tunability under external fields, making it an ideal candidate for exploring advanced valley control mechanisms ~\cite{li2022two,luo2020valley}.
\begin{figure}[htp]
	\includegraphics[width=0.5\textwidth]{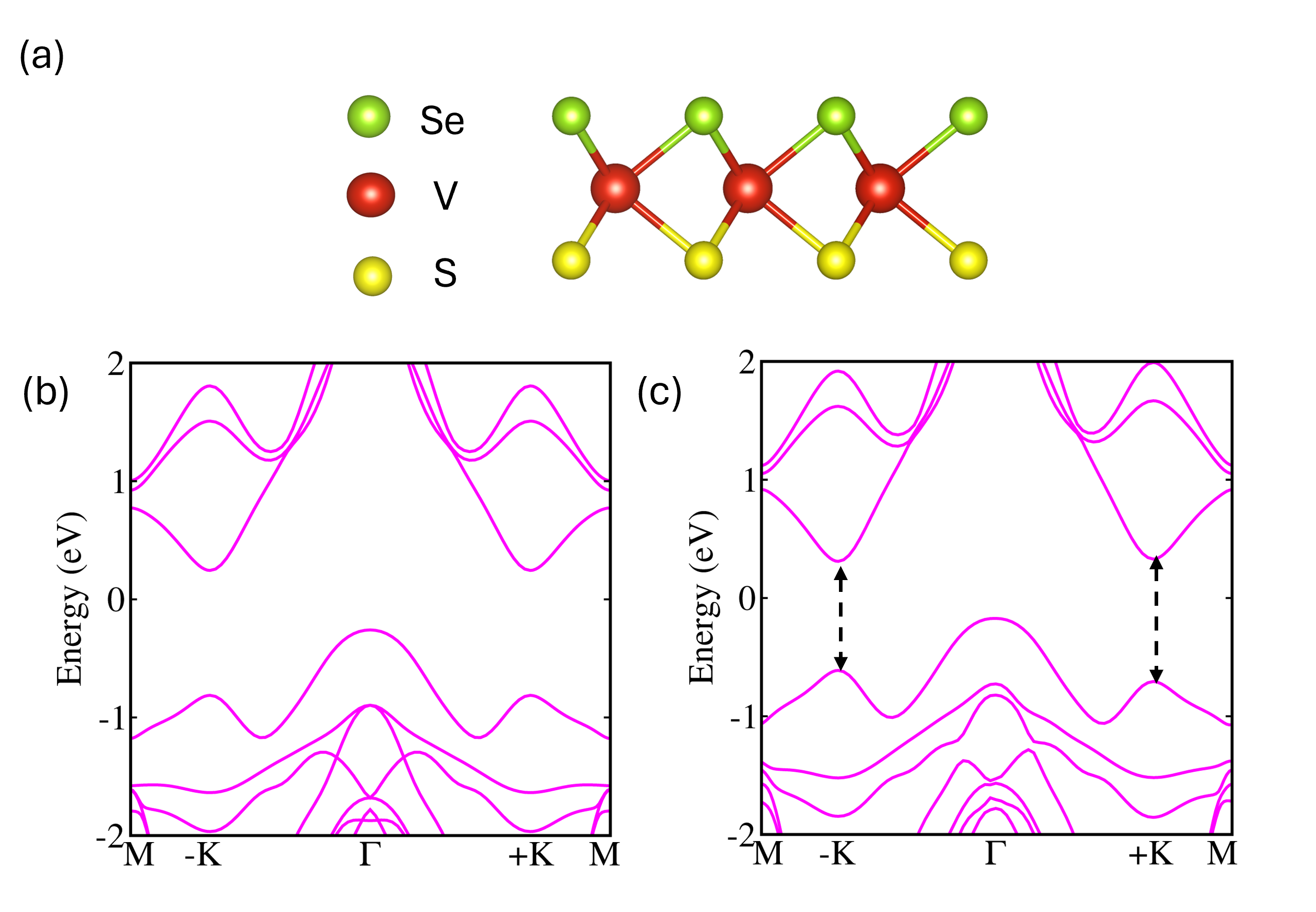}
	\caption{(a) Crystal structures of monolayer VSSe. (b) Band structures of monolayer VSSe without SOC. (c) Band structures of monolayer VSSe with SOC. The Fermi level is set to zero.}
	\label{5-1}
\end{figure}

The crystal structure and electronic band structure of VSSe is shown in Fig. \ref{5-1}(a) and (b), repectively. As depicted in its band structure Fig. \ref{5-1}(c), the coexistence of SOC and inherent exchange interaction in monolayer VSSe lifts the energy degeneracy between these two valleys, leading to valley polarization. When an external in-plane electric field is applied, Bloch electrons from either the $+K$ and $-K$ valley acquire transverse velocities. This occurs under the influence of an effective magnetic field generated by the Berry curvature in that respective valley, which in turn gives rise to the AVH effect. While the valley polarization in monolayer VSSe can be tuned by an external magnetic field, thereby controlling the sign of the AVH effect. However the magnetic approach is generally energy-intensive. Therefore, an alternative, reversible, and nonvolatile electrical method for manipulating valley polarization is highly desirable.
\begin{figure}[htp]
	\includegraphics[width=0.5\textwidth]{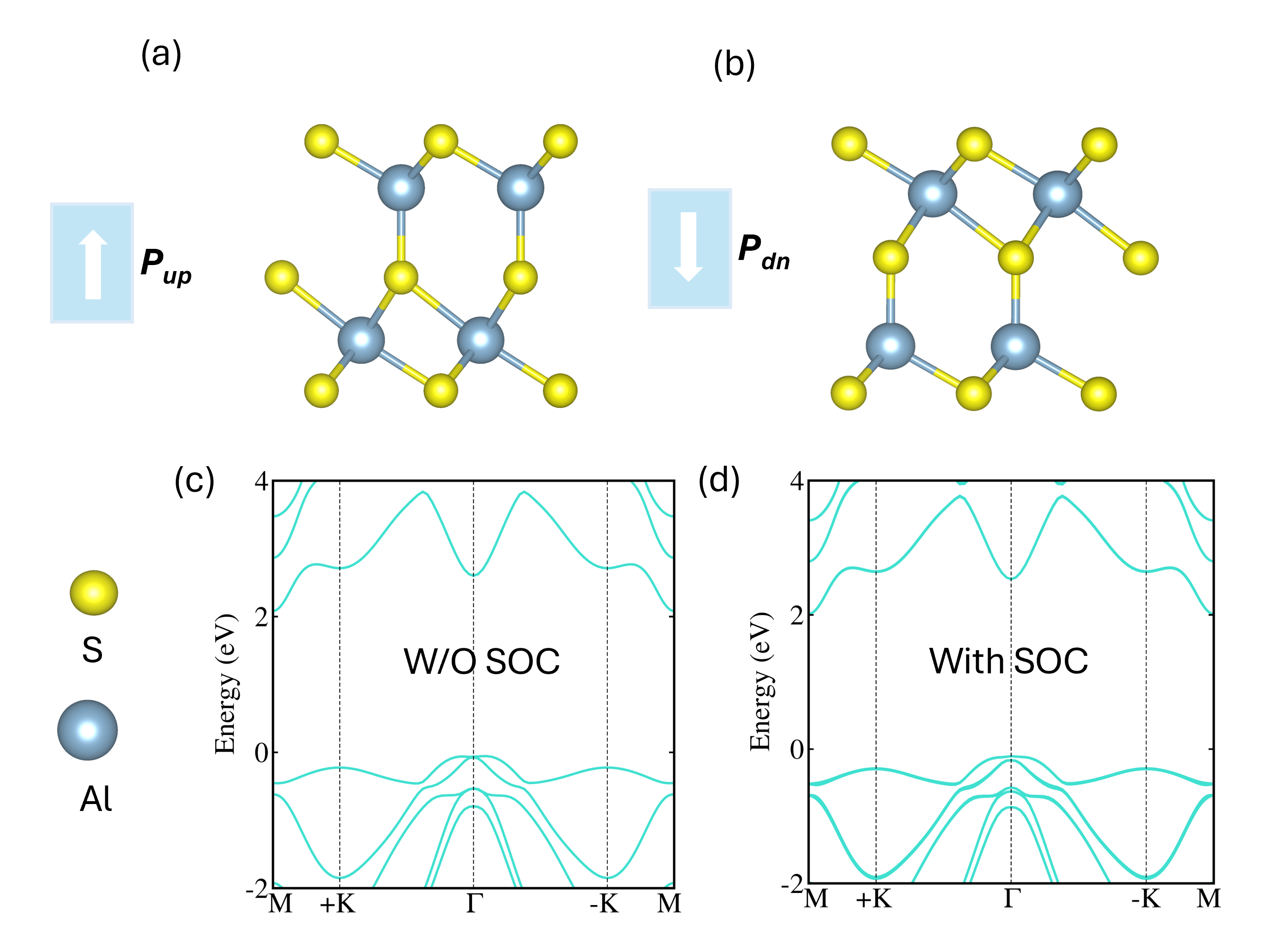}
	\caption{(a), (b) Crystal structures of monolayer Al$_2$S$_3$ from side views with different ferroelectric polarization. (c) Band structures of monolayer Al$_2$S$_3$ without SOC. (d) Band structures of monolayer Al$_2$S$_3$ with SOC. The Fermi level is set to zero.}
	\label{5-2}
	
\end{figure}

\begin{figure*}[htp]	
	\begin{center}
		\includegraphics[width=0.7\textwidth]{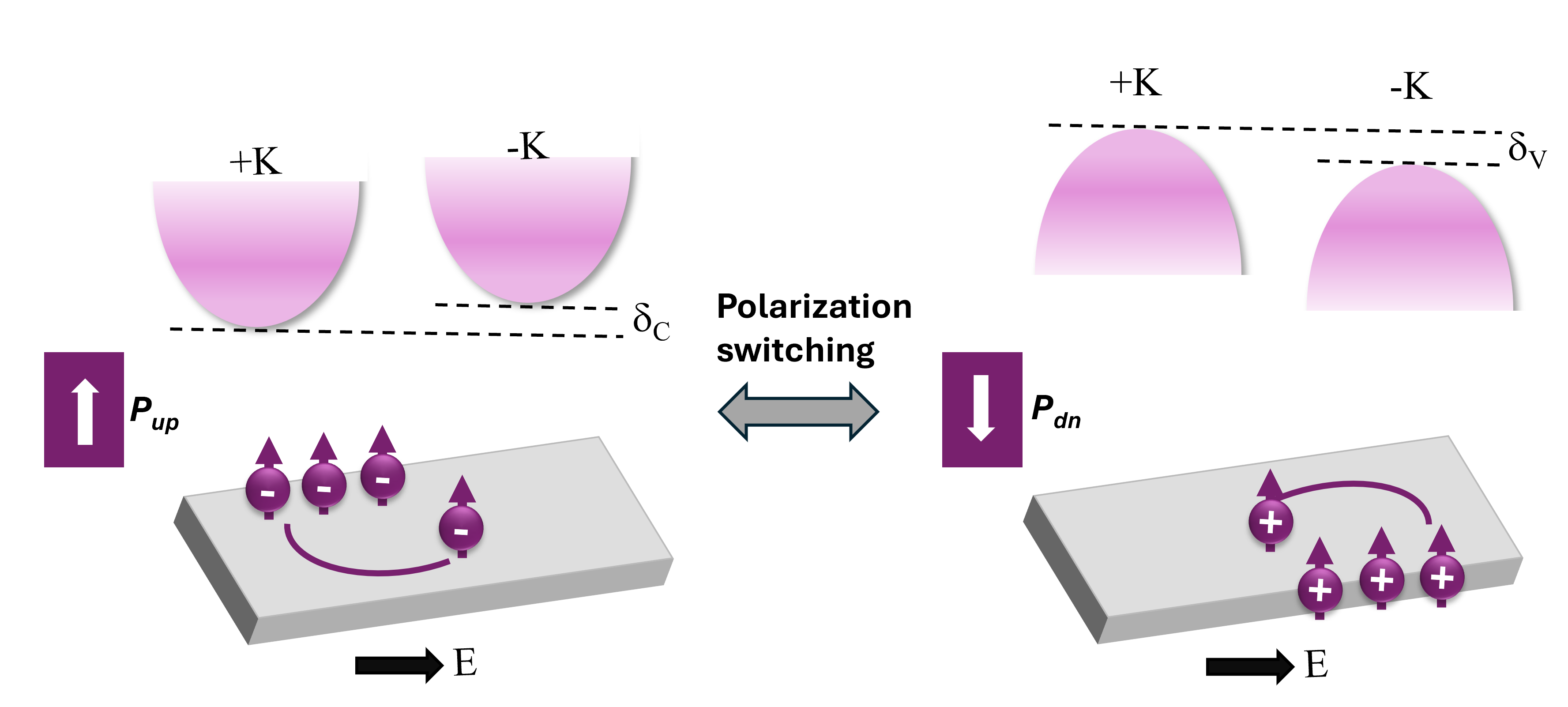}
		\caption{Design principle based on multiferroic heterostructure. Diagrams of band structures around the $+K$ and $-K$ valleys and FE-controlled AVH effect under in-plane electric field.}
		\label{5-3}
	\end{center}
\end{figure*}

To achieve this electrical control, 2D ferroelectric materials emerge as ideal candidates. These materials possess two stable, electrically switchable polarized states that are nonvolatile, meaning the chosen polarization direction persists even after the external electrical stimulus is removed. For this we chose 2D Al$_2$S$_3$ as a ferroelectric substrate. This material's crystal structure, with its distinct polarization states, is detailed in Fig. \ref{5-2}(a) and (b). The stable 2D Al$_2$S$_3$ monolayer features a quintuple-layer structure, specifically S-Al-S-Al-S, characterized by the  \textit{P}3\textit{m}1 space group. Our calculated lattice parameter of 3.59 \AA\ aligns well with previous findings ~\cite{fu2018intrinsic}. The material's ferroelectricity stems from the asymmetric displacement of atoms in its central sulfur layer, enabling reversible spontaneous electric polarization in out-of-plane orientations. We denote these ferroelectric polarization states as P$_{up}$ and P$_{dn}$, corresponding to the middle sulfur atom's vertical alignment with the upper and lower aluminum layers, respectively. Al$_2$S$_3$ is part of the well-established In$_2$Se$_3$ family of ferroelectrics, which are known to exhibit room-temperature ferroelectric properties experimentally ~\cite{ding2017prediction, xue2018room}. Our analysis of the band structure for 2D Al$_2$S$_3$ as shown in Fig. \ref{5-2}(a) confirms its semiconductor nature with an indirect band gap of 2.97 eV. Here, the valence band maximum (VBM) is situated along the $K$-$\Gamma$ line of the Brillouin zone, while the conduction band minimum (CBM) is at the $\Gamma$ point. While we also considered the impact of SOC, it introduced only minor modifications to the overall band structure as shown in Fig. \ref{5-2}(c) and (d).
Placing a ferrovalley VSSe monolayer on a ferroelectric substrate allows for electrically controlled modulation of both valley polarization and the AVH effect. Our design principle, visually represented in Fig. \ref{5-3}, demonstrates that reversing the ferroelectric substrate's polarization can switch the VSSe valley polarization from the VBM to the CBM. This transition can even submerge the VBM or CBM valley into trivial bands, a distinct feature not achievable through doping. Crucially, this ferroelectric substrate provides reversible and nonvolatile tuning of the AVH effect's sign, a capability doping alone cannot match.
\begin{figure*}[htp]	
	\begin{center}
		\includegraphics[width=1\textwidth]{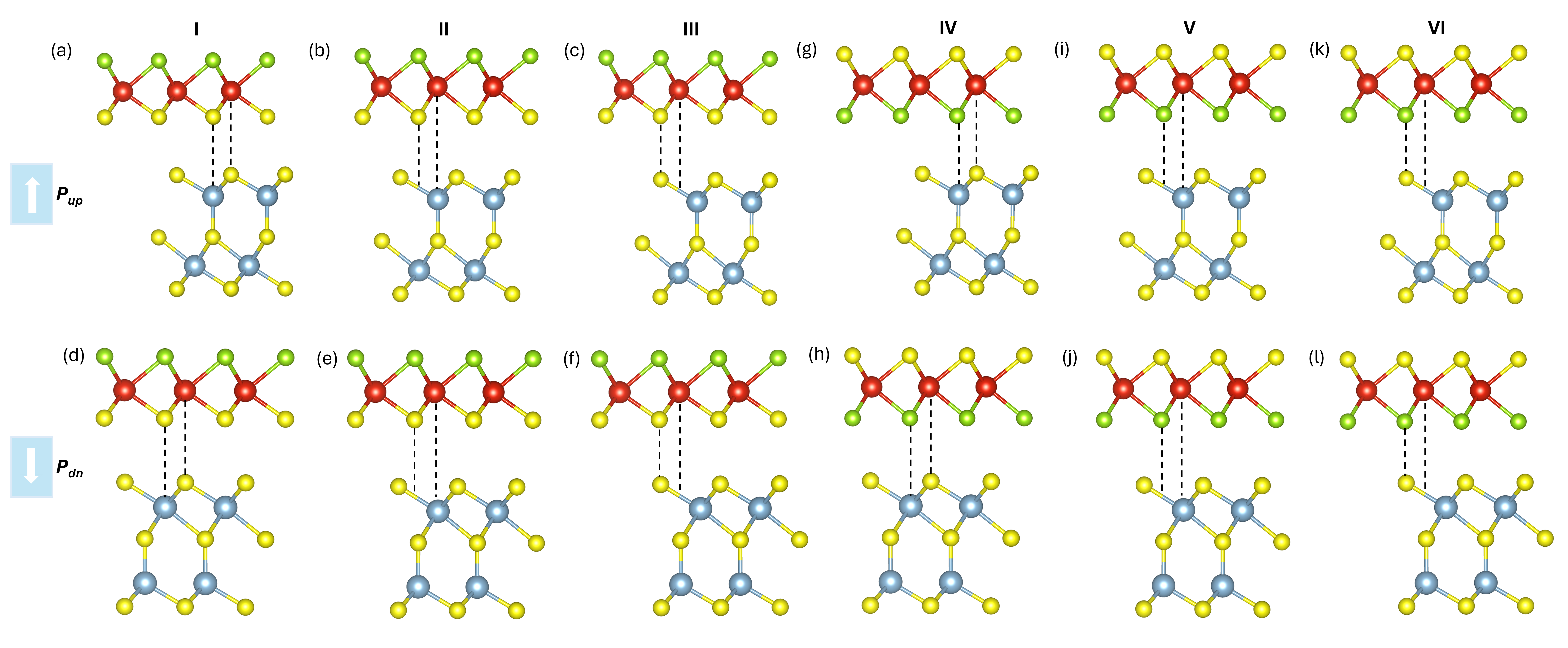}
		\caption{Crystal structures of different stacking configurations of the VSSe/Al$_2$S$_3$ heterostructures with (a), (c), (e), (g), (i), (k) P$_{up}$ and (b), (d), (f), (h), (j), (l) P$_{dn}$ polarization.}
		\label{5-4}
	\end{center}
\end{figure*}

\begin{table}[h!]
	\caption {Energy of different stacking configurations of VSSe/Al$_2$S$_3$ heterostructure in units of eV.}
	\begin{center}
		\setlength\extrarowheight{+4pt}
		\begin{tabular}[c]{|c|c|c|} \hline		
			Configuration-VSSe/Al$_2$S$_3$& P$_{up}$ & P$_{dn}$ \\ \hline
			I & 0& 0     \\ \hline
			II & 0.250 & 0.232     \\ \hline
			III & 0.255 &  0.234    \\ \hline
			IV & 0.342 &  0.333    \\ \hline
			V & 0.345 & 0.334     \\ \hline
			VI & 0.344 & 0.330     \\ \hline
		\end{tabular}		
		\label{T1}
	\end{center}
\end{table}

To engineer the VSSe/Al$_2$S$_3$ multiferroic heterostructure, we aligned the unit cells of 2D Al$_2$S$_3$ and 2D VSe$_2$. The minimal lattice mismatch of 3.6\% between these materials suggests that experimental fabrication of this heterostructure should be readily achievable. We investigated six potential stacking configurations for both polarization states (P$_{up}$ and P$_{dn}$) each, as shown in Fig. \ref{5-4}. Our calculations of the relative energies for these various structures (Table \ref{T1}) indicate that the type-I configuration is energetically the most stable for both polarization directions. Susequently, we have calculated the energies of type-I configuration in different magnetic states namely FM, AFM1, AFM2, and AFM3. The up and down arrows represent spin-up and spin-down Vanadium atoms, respectively as shown in section-I of Supplemental Materials. Crucially, the ferromagnetic (FM) ground state of VSSe is preserved within this multiferroic heterostructure, as confirmed by the energies of various magnetic configurations listed in Table shown in section-I of SM. All the energies has been calculated using HSE06 functional.
\begin{figure}[htp]
	\includegraphics[width=0.5\textwidth]{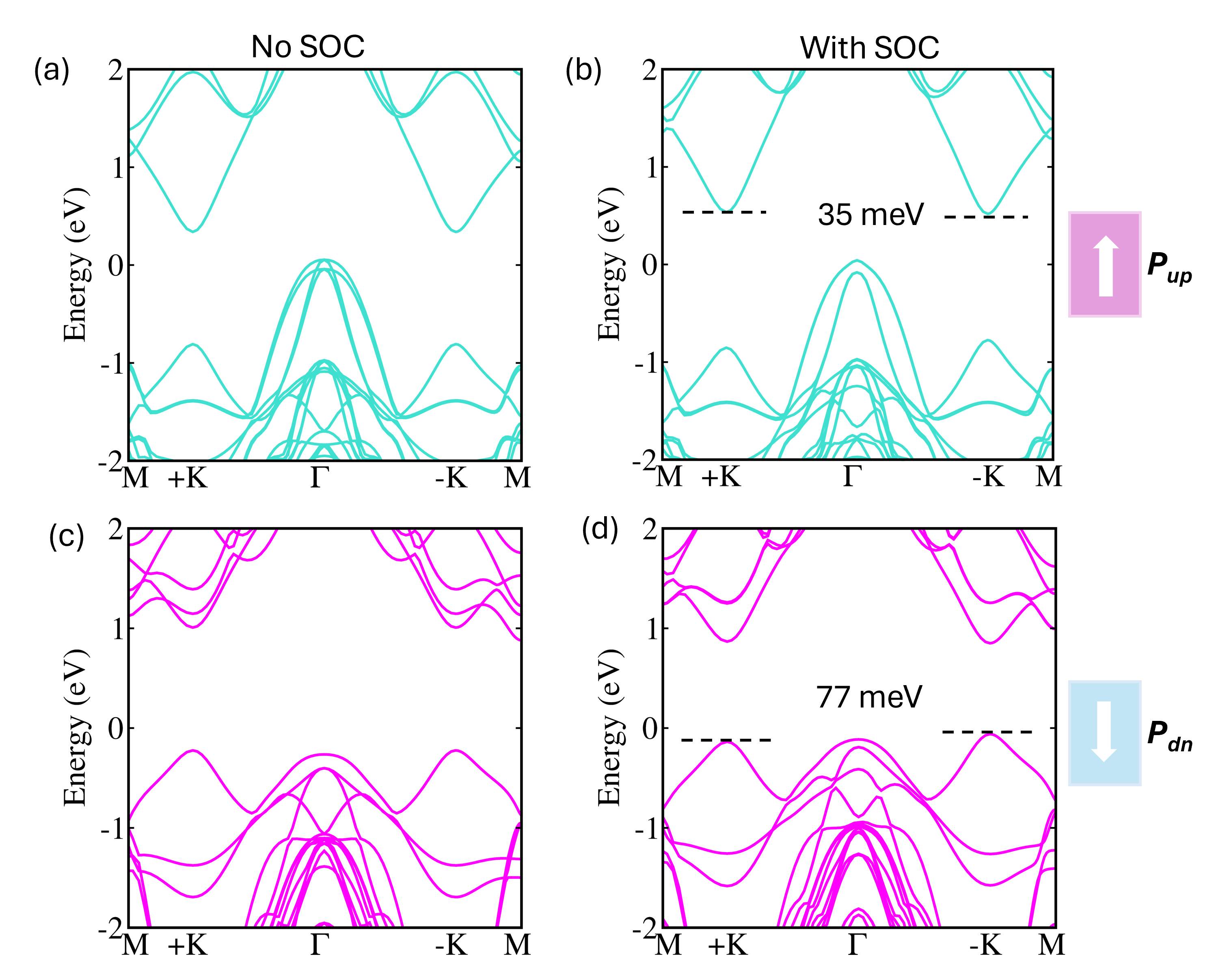}
	\caption{Band structures of (a) type-I P$_{up}$ and (c) type-I P$_{dn}$ configurations without SOC. Band structures of (b) type-I P$_{up}$ and (d) type-I P$_{dn}$  configurations with SOC. White arrows show the FE polarization direction. The Fermi level is set to zero.}
	\label{5-6}
\end{figure}
Further, the electronic band structures for type-I P$_{up}$ is displayed in Fig. \ref{5-6}(a). The energy band structure shows an indirect band-gap semiconductor with 0.29 eV band gap. Here, the VBM is located at the $\Gamma$ point, while the CBM resides at the $+$K/$-$K  point. This arrangement creates a typical staggered-gap (type-II) band alignment, where the VBM is primarily contributed by Al$_2$S$_3$ and the CBM by the spin-up channel of VSSe. Moreover, both the conduction and valence bands have degenerate valleys at $+$K and $-$K points. The valleys of the valence band are submerged within the trivial bands of ferroelectric substrate. In contrast, the type-I P$_{dn}$ configuration undergoes a significant transformation in its band structure. Its indirect band gap expands to 0.88 eV, notably wider than that of the type-I P$_{up}$ state. As depicted in Fig. \ref{5-6}(c), the VBM shifts to the $+$K/$-$K point, while the CBM moves to the M point. Although the type-II band alignment is maintained, the dominant contributions to the band edges are effectively inverted compared to the type-I P$_{up}$ configuration. Specifically, the VBM is now primarily governed by the spin-up channel of VSSe, with the CBM being dominated by Al$_2$S$_3$. Crucially, unlike the type-I P$_{up}$ case, the valleys within the VBM remain well-preserved at the $+$K and $-$K points, whereas the valleys in the lowest conduction band become submerged.

When we include the SOC effect, as illustrated in Fig. \ref{5-6}(b) and (d), valleys become non-degenerate, resulting in spontaneous valley polarization. We calculated the spontaneous valley polarization for type-I P$_{up}$ case in the CBM to be $\Delta$E$_c$ = 35 meV, and for type-I P$_{dn}$ in the VBM, to be $\Delta$E$_v$ = 77 meV, which is a substantial. These values significantly surpass those found in some experimentally demonstrated magnetic proximity systems (0.3-1.0 meV)~\cite{seyler2018valley} and existing ferrovalley materials~\cite{he2021two}. Such robust polarization can effectively resist the annihilation of valley polarization caused by kinetic energy near room temperature (25 meV). It is important to note that in the type-I P$_{up}$ heterostructure, the VSSe valleys in the VBM become submerged in the trivial bands of Al$_2$S$_3$, which limit their potential utilization. Conversely, for the type-I P$_{dn}$ configuration, the VSSe valleys in the CBM are similarly submerged. Consequently, the effective response driven by valley polarization switches from the CBM to the VBM simply by flipping the ferroelectric polarization. This transition is accompanied by a fundamental change in valley physics and the AVH effect. This means the AVH effect becomes highly controllable by reversing the electric polarization of Al$_2$S$_3$, which greatly benefits the development of reversible and nonvolatile controllable valleytronic devices. This approach stands apart from heterostructures like VSe$_2$/Sc$_2$CO$_2$~\cite{lei2021nonvolatile}, where ferroelectricity primarily enables only an on-off switching of the AVH effect. Furthermore, a detailed examination of the spin-polarized electronic band structures, as presented in section-II of SM, reveals distinct spin contributions for both type-I P$_{up}$ and type-I P$_{\text{dn}}$ configurations. These plots indicate that in the type-I P$_{up}$ configuration, valleys near the Fermi level in conduction band are primarily contributed by spin-up electrons of Vanadium. Conversely, in type-I P$_{\text{dn}}$ configuration, the dominant contribution near the Fermi level in the valence band comes from the spin-up holes of Vanadium.  

\begin{figure}[htp]
	\includegraphics[width=0.5\textwidth]{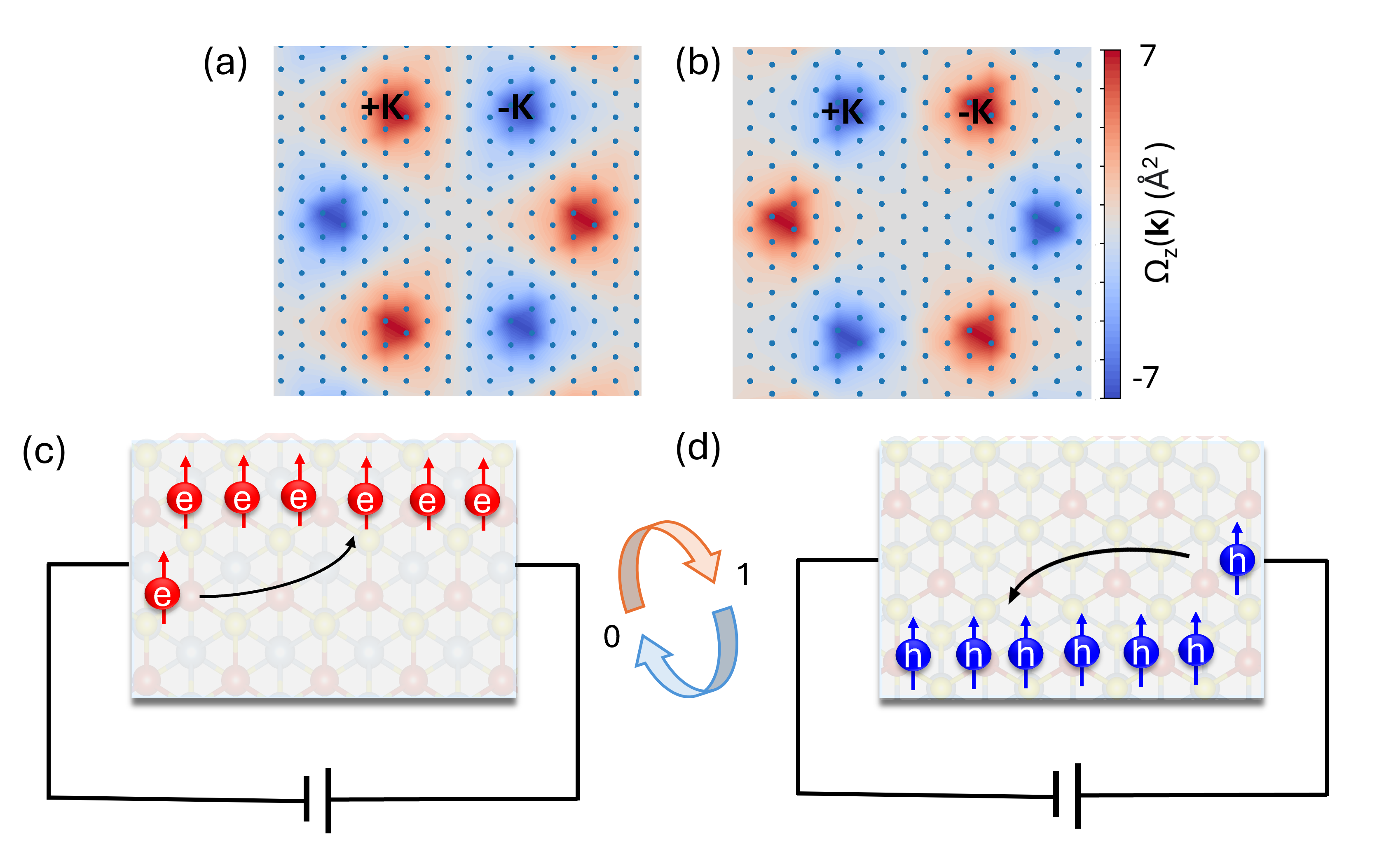}
	\caption{Contour maps of the Berry curvatures across the 2D Brillouin zone for the (a) type-I P$_{up}$ and (b) type-I P$_{dn}$ configurations. (c) Schematic of the AVH effect and memory device in (c) type-I P$_{up}$ and (d) type-I P$_{dn}$ configurations.}
	\label{5-8}
\end{figure}

Furthermore, Berry curvature has been an invaluable tool for examining the Hall effect. Particularly in systems where space inversion symmetry is broken, it has been observed that charge carriers within the $+$K and $-$K valleys tend to acquire a non-zero Berry curvature. Consequently, when time-reversal symmetry is also broken, an interesting characteristic emerges: a contrasting feature between valleys becomes apparent. Therefore, to understand the valley-related properties of heterostructure we have calculated the Berry curvature using Kubo formula ~\cite{thouless1982quantized}: 
\begin{equation}
\Omega (\mathrm{K}) = - \sum_{n} \sum_{m \neq n} f_n \frac{2 Im \langle \psi_{nk} | v_x | \psi_{mk} \rangle  \langle \psi_{mk} | v_y | \psi_{nk} \rangle} {(E_n - E_m)^2}
\end{equation}
Here, \(f_n\) is the Fermi-Dirac distribution function, \(E_{n}\) represents the eigenvalue of the Bloch state \(|\psi_{nk}\rangle\) and \(\hat{v}_x\)/\(\hat{v}_y\) are the velocity operator components. As depicted in Fig. \ref{5-8}(a) and (b), the Berry curvatures around the $+K$ and $-K$ valleys exhibit distinct absolute values and opposite signs, a clear indication of a robust valley-contrasting feature. A key finding is the ability to switch the valley polarization from the CBM to the VBM simply by reversing the ferroelectric polarization of the Al$_2$S$_3$ substrate. This phenomenon occurs because the time-reversal operator, which transforms the orbital part of the Bloch function to its complex conjugate and flips the spin, causes the VBM and CBM at both the $+K$ and $-K$ points to display Berry curvatures with opposing signs. Consequently, we observe a complete reversal of the Berry curvature signs in response to ferroelectric polarization switching.


Since, the anomalous Hall conductivity is directly proportional to the integral of the Berry curvature over the Brillouin zone, the net Hall currents generated in the type-I$_{up}$ and type-I$_{dn}$ states are clearly distinguishable. The ability to switch valley physics between the conduction and valence bands via ferroelectricity allows for an on-off control in both electron-doped and hole-doped scenarios. This demonstrates that the anomalous valley Hall effect is significantly realized and effectively controlled through ferroelectric switching.

These findings highlight the promising potential of the VSSe/Al$_2$S$_3$ heterostructure for designing valleytronic memory devices. We have developed a prototype device based on this heterostructure, depicted in Fig. \ref{5-8}(c) and (d). In this prototype, the electric polarization in the Al$_2$S$_3$ is toggled by a vertical external voltage. When the ferroelectric substrate's polarization is in the P$_{up}$ configuration (Fig. \ref{5-8}(c)), spin-up electrons accumulate on the up side of the VSSe monolayer due to valley polarization at the CBM, yielding a measurable transverse Hall voltage. We define this as the "0" state of the device. Conversely, when the ferroelectric substrate's polarization direction is reversed to P$_{dn}$ (Fig. \ref{5-8}(d)), spin-up holes become the majority carriers, resulting in a different transverse Hall voltage signal strength, which we identify as the "1" state. This design allows for a damage-free data reading process by simply checking the differences in the ferroelectric coupling-induced Hall voltage signal strength, effectively avoiding the destructive effects often associated with directly sensing polarized states in memory devices.

\section{CONCLUSIONS}
We have theoretically developed a design principle for achieving reversible and nonvolatile control of the AVH effect. Our first-principles calculations confirm that stacking ferrovalley VSSe on a ferroelectric Al$_2$S$_3$ monolayer allows for effective tuning of the AVH effect. We can electrically control the VSSe valley polarization, switching it from the VBM to the CBM by simply reversing the electric polarization of ferroelectric substrate. The electrical control offers a highly desirable, energy-efficient alternative to the more intensive magnetic methods typically used for controlling valley polarization. Critically, the sign of transverse Hall voltage is directly controlled by the ferroelectric polarization. This ferroelectric-driven and tunable AVH effect in such a multiferroic heterostructure presents a promising avenue for all-electric data memory devices.

\section*{ACKNOWLEDGEMENT}
	A.P. acknowledges IIT Delhi for the senior research fellowship. S.B. acknowledges financial support from SERB under a core research grant (grant no. CRG/2019/000647) to set up his High Performance Computing (HPC) facility ‘‘\textit{Veena}’’ at IIT Delhi for computational resources.

	


\end{document}